\documentclass[aps,prb,twocolumn, superscriptaddress, showpacs, a4paper, 10pt]{revtex4-1}
\usepackage [dvips]{graphicx}
\usepackage{graphicx,color}
\usepackage{ulem}
\usepackage{amsmath}
\usepackage{bm}
\usepackage{dcolumn}
\usepackage{multirow}
\usepackage{booktabs}
\usepackage[
	bookmarks=true,
	colorlinks=true,
	urlcolor=blue,
	linkcolor=blue,
	citecolor=blue,
	pdfpagemode=UseNone,
	pdfstartview=FitH,
	pdfencoding=auto
	]{hyperref}
\usepackage[all]{hypcap}
\begin{document}

\title{Strong out-of-plane magnetic anisotropy of Fe adatoms on Bi$_2$Te$_3$}

\author{T.~Eelbo}
\email{teelbo@physnet.uni-hamburg.de}
\affiliation{Institute of Applied Physics, University of Hamburg, Jungiusstr. 11, 20355 Hamburg, Germany}

\author{M.~Wa\'sniowska}
\email{mwasniow@physnet.uni-hamburg.de}
\affiliation{Institute of Applied Physics, University of Hamburg, Jungiusstr. 11, 20355 Hamburg, Germany}

\author{M.~Sikora}
\affiliation{AGH University of Science and Technology, Academic Centre for Materials and Nanotechnology, al. Mickiewicza 30, 30-059 Krak\'{o}w, Poland }

\author{M.~Dobrza\'nski}
\affiliation{AGH University of Science and Technology, Faculty of Physics and Applied Computer Science, al. Mickiewicza 30, 30-059 Krak\'{o}w, Poland}

\author{A.~Koz{\l}owski}
\affiliation{AGH University of Science and Technology, Faculty of Physics and Applied Computer Science, al. Mickiewicza 30, 30-059 Krak\'{o}w, Poland}

\author{A.~Pulkin}
\affiliation{Institute of Theoretical Physics, Ecole Polytechnique F{\'{e}}d{\'{e}}rale de Lausanne (EPFL), CH-1015 Lausanne, Switzerland}

\author{G.~Aut\`{e}s}
\affiliation{Institute of Theoretical Physics, Ecole Polytechnique F{\'{e}}d{\'{e}}rale de Lausanne (EPFL), CH-1015 Lausanne, Switzerland}

\author{I.~Miotkowski}
\affiliation{Department of Physics, Purdue University, 525 Northwestern Avenue, West Lafayette, IN 47907, USA}

\author{O.~V.~Yazyev}
\affiliation{Institute of Theoretical Physics, Ecole Polytechnique F{\'{e}}d{\'{e}}rale de Lausanne (EPFL), CH-1015 Lausanne, Switzerland}

\author{R.~Wiesendanger}
\affiliation{Institute of Applied Physics, University of Hamburg, Jungiusstr. 11, 20355 Hamburg, Germany} 

\date{\today}

\begin{abstract}

The electronic and magnetic properties of individual Fe atoms adsorbed on the surface of the topological insulator Bi$_2$Te$_3$(111) are investigated. Scanning tunneling microscopy and spectroscopy prove the existence of two distinct types of Fe species, while our first-principles calculations assign them to Fe adatoms in the hcp and fcc hollow sites. The combination of x-ray magnetic circular dichroism measurements and angular dependent magnetization curves reveals out-of-plane anisotropies for both species with anisotropy constants of $K_{\text{fcc}} = (10 \pm 4)$~meV/atom and $K_{\text{hcp}} = (8 \pm 4)$~meV/atom. These values are well in line with the results of calculations.

\end{abstract}


\maketitle

In recent years a new class of materials, the so-called topological insulators (TIs) realizing a novel topologically non-trivial electronic phase driven by strong spin-orbit interactions, has attracted considerable attention~\cite{Moore2010,Hasan2010,Qi2011}. These materials are expected to offer the possibility of the experimental realization of new physical phenomena and functionalities like the Majorana fermions~\cite{Fu2008}, the quantum spin Hall effect in zero external magnetic fields~\cite{Bernevig2006,Koenig2007}, or an image magnetic monopole induced by a charge near the TI surface via the topological magneto-electric effect~\cite{Qi2009,Zang2010}. Angle-resolved photoemission spectroscopy (ARPES) measurements reveal the presence of topologically protected metallic surface states within the bulk band gap of TIs such as Bi$_2$Se$_3$ and Bi$_2$Te$_3$~\cite{Xia2009,Chen2009}. 

Importantly, doping by external magnetic impurities strongly modifies the electronic structure of pristine TIs resulting in novel topological electronic phases~\cite{Chen2010,Hor2010,Wray2011,Okada2011,Chang2013}. In previous studies performed on Bi$_2$Te$_3$ and Bi$_2$Se$_3$ doped with Mn and Fe magnetic impurities ferromagnetic ordering of the impurities below 12~K and 2~K, respectively, was discovered~\cite{Hor2010,Chen2010,Salman2012}. For diluted magnetic dopants exchange-coupled to the helical electron gas on the surface of TIs, a magnetic out-of-plane anisotropy was predicted~\cite{Abanin2011,Nunez2012} and subsequently confirmed in recent experiments for Mn dopants in Bi$_2$Te$_3$ and Bi$_2$Se$_3$~\cite{Hor2010,Xu2012}. However, theory and experiment disagree in the case of Fe adatoms on Bi$_2$Se$_3$. While theory predicted an out-of-plane anisotropy as well~\cite{Li2012}, x-ray magnetic circular dichroism measurements proved an easy-plane anisotropy of the Fe adatoms~\cite{Honolka2012}. For this reason, the complexity of interaction between magnetic impurities and TI surfaces calls for further investigations of magnetically doped TIs.

Here, we report a combined experimental and theoretical investigation on Fe atoms adsorbed on the Bi$_{2}$Te$_{3}$(111) surface by means of scanning tunneling microscopy/spectroscopy (STM/STS), x-ray absorption spectroscopy (XAS), x-ray magnetic circular dichroism (XMCD), and first-principles calculations. Two distinct configurations of Fe adatoms have been observed in the STM/STS experiments while our first-principles calculations help in assigning them to fcc and hcp hollow sites. In addition, the angular dependence of element specific magnetization profiles reveals strong out-of-plane magnetic anisotropies as predicted by first-principles calculations.

All experiments were carried out in ultrahigh vacuum (UHV) systems. The STM/STS measurements were performed at 5~K with a base pressure below 10$^{-10}$~mbar in an experimental setup described elsewhere~\cite{Meckler2009}. The single crystals were cleaved \textit{in situ} at low temperatures. Fe was deposited at 12~K in minute amounts using an electron-beam evaporator. The differential conductance (d$I$/d$U$) was measured via lock-in technique with a modulation voltage of 20~mV and a frequency of 5~kHz. The XAS and XMCD measurements were performed in the total-electron-yield mode at the ID08 beamline of the European Synchrotron Radiation Facility using almost 100\% circularly polarized light. After cleaving the samples inside an UHV at room temperature they were immediately cooled down to 6~K within the next 15 min. Fe was again deposited by means of an electron-beam evaporator with the substrate held at $T=10$~K. Magnetic fields up to 5~T were applied collinear with the incident x-ray beam to magnetize the sample. Out-of-plane and in-plane properties were investigated by rotating the sample from 0$^{\circ}$ (normal incidence) to 70$^{\circ}$ (grazing incidence). All spectra were normalized with respect to the incident beam intensity and the XAS pre-edge intensity at the Fe $L_3$ edge at 705~eV.

\begin{figure}
\begin{center}
\includegraphics[width=0.30\textwidth]{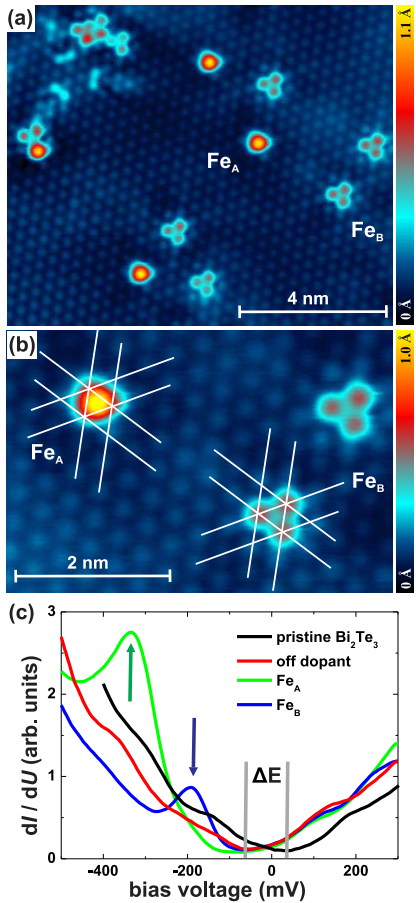}
\caption{(Color online) (a) STM constant current map of the Bi$_{2}$Te$_{3}$ surface covered by 0.01 MLE of Fe adatoms measured at $U=-0.4$~V and $I=0.3$~nA. (b) Atomically resolved topography showing two different Fe adatoms (Fe$_{\rm{A}}$ and  Fe$_{\rm{B}}$) occupying different adsorption sites ($U=-0.4$~V and $I=0.3$~nA). (c) STS measurements performed on Bi$_2$Te$_3$ before and after Fe deposition. The ``off dopant'' spectrum was taken after deposition far away from any adsorbate. The spectra were acquired at $U=0.3$~V and $I=0.2$~nA.
}\label{Fig1}
\end{center}
\end{figure}

Figure~\ref{Fig1}(a) presents a STM image of the (111) surface of Bi$_2$Te$_3$ after the deposition of 0.01 monolayer equivalent (MLE) of Fe adatoms. Two different types of Fe atoms can be easily distinguished, labeled Fe$_{\rm{A}}$ and Fe$_{\rm{B}}$. The two species have different apparent heights and shapes that strongly depend on the applied bias voltage as well as their lateral position with respect to the underlying substrate. The atomically resolved protrusions of the surface are caused by the topmost Te layer of the Bi$_2$Te$_3$ quintuple layer. We assign Fe$_{\rm{A}}$ and Fe$_{\rm{B}}$ to different hollow sites [cf. Fig.~\ref{Fig1}(b)]. Independent of the bias voltage, Fe$_{\rm{A}}$ shows an apparent height larger than that of Fe$_{\rm{B}}$. Fe$_{\rm{A}}$ appear as triangle-shaped features pointing to the upper left [in Fig.~\ref{Fig1}(b)] with a central protrusion independent of the polarity. In contrast, Fe$_{\rm{B}}$ is pointing to the lower right (at negative bias voltages) and shows up as a combination of three minor protrusions. Note, that Fe$_{\rm{B}}$ is caused by a single Fe impurity, which was verified by a control experiment depositing the same amount of Fe on a W(110) substrate and comparing the density of adsorbates on both surfaces. At positive voltages the shape of Fe$_{\rm{B}}$ varies strongly; i.e., it is spatially enlarged and shows a different apparent shape. The observation of two different types of Fe species is contrary to a recent report on the particular system of Fe/Bi$_2$Te$_3$(111)~\cite{West2012}. West \textit{et al.} found the existence of only a single species of adatoms, whereas these different findings are most likely caused by the different preparation conditions since West \textit{et al.} deposited and investigated the Fe adatoms at higher temperatures (50~K). In this case the thermal energy might be sufficient to overcome the diffusion barrier between both adsorption sites which would consequently result in the observation of only a single type of Fe adatoms. We note, that in the present study the Fe atoms do not appear as dark triangular depressions as is the case for diluted Fe dopants inside Bi$_2$Te$_3$ after room-temperature annealing~\cite{West2012}, confirming that in the present study Fe atoms do not substitute Bi atoms. Concerning the relative abundance, we find a ratio of approximately $60 \%$~to~$40\%$ in favor of Fe$_{\rm{B}}$. In addition, the electronic properties of the Fe adatoms were investigated by means of STS [Fig.~\ref{Fig1}(c)]. Therefore, spectra of the pristine Bi$_2$Te$_3$ prior to Fe deposition were taken first. The spectra of the Fe-coated substrate are shifted by $\approx-90$~mV compared to the spectra obtained on pristine Bi$_2$Te$_3$. We note, that in the case of Bi$_2$Te$_3$, the global minimum within the spectra cannot be straightforwardly assigned to the Dirac point (DP) because the DP is buried by the onset of the bulk valence band~\cite{Alpichshev2010}. Nevertheless, the shift to lower energy indicates the transfer of electrons from the Fe impurities. This indicates effective $n$-doping of the substrate in agreement with previous reports~\cite{Chen2010,Honolka2012,Song2012}. Furthermore, we find different electronic signatures for Fe$_{\rm{A}}$ and Fe$_{\rm{B}}$. While Fe$_{\rm{A}}$ shows an additional peak at $\approx-335$~mV we find a peak at $\approx-190$~mV in the case of Fe$_{\rm{B}}$. Moreover, by means of STS we cannot detect a gap-opening caused by the Fe deposition, neither in the spectra acquired on both types of adatoms nor at close or far distances.

To clarify the details of the atomic structure and the potential energy surface of Fe adatoms on Bi$_2$Te$_3$ we carried out first-principles theoretical investigations. The calculations were performed within the density functional theory (DFT) framework employing the local density approximation including an on-site Coulomb repulsion term (LDA+$U$) on the Fe atoms ($U=2.2$~eV~\cite{Cococcioni2005}). The latter is indispensable for correctly reproducing both the ground-state magnetic moment and the magnetic anisotropy energies of transition metal adatoms~\cite{Donati2013}. Spin-orbit effects were included by using fully relativistic pseudopotentials acting on valence electron wavefunctions represented in the two-component spinor form \cite{Corso2005}. The fcc-to-hcp transition path was found using the nudged elastic band approach~\cite{Henkelman2000}. We used the \textsc{quantum-espresso} package~\cite{Giannozzi2009} for performing our calculations.

\begin{figure}
\begin{center}
\includegraphics[width=0.35\textwidth]{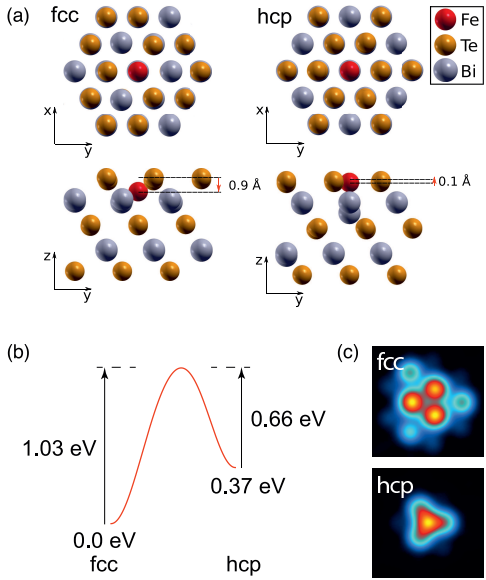}
\caption{(Color online) (a) Atomic configurations of Fe atoms adsorbed in the fcc and hcp adsorption sites as calculated from first principles. The displacements of the Fe adatoms with respect to the Te top layer are indicated. (b) Schematic drawing of the potential energy surface of the Fe adatoms in fcc and hcp sites on the Bi$_2$Te$_3$ surface calculated from first principles. (c) Simulated STM images of Fe adatoms in the fcc and hcp sites for a bias voltage of $U=-0.4$~V.}\label{Fig2}
\end{center}
\end{figure}

Two possible adsorption sites of the Fe adatoms were considered: the fcc hollow site and the hcp hollow site [Fig.~\ref{Fig2}(a)]. By means of the LDA+$U$ approach, the fcc site was found to be lower in energy by 0.37~eV [Fig.~\ref{Fig2}(b)]. Its atomic configuration is characterized by a strong relaxation ($\sim$0.9~\AA) into the surface [Fig.~\ref{Fig2}(a)]. This finding contrasts with the case of Bi$_2$Se$_3$, where calculations utilizing the general gradient approximation give binding energies of the Fe adatoms in both positions differing by only 0.07~eV~\cite{Honolka2012}. The difference can be assigned to a larger lattice constant of Bi$_2$Te$_3$ which allows for a larger displacement of the fcc-site Fe adatom. Thus, the observation of two different Fe species with almost equal populations points to nonequilibrium thermodynamics. This conclusion is further strengthened by the calculated diffusion barrier of 0.66~eV for the transition from the hcp site to the energetically more favorable fcc site [Fig.~\ref{Fig2}(b)]. Since the deposition and measurements were performed at low temperatures, Fe adatoms on the surface of Bi$_2$Te$_3$ can be assumed to be immobile, and hence, both adsorption sites are expected to be populated. Further insight is obtained from the simulated STM images of adatoms in both adsorption sites, shown in Fig.~\ref{Fig2}(c), which clearly resemble the experimental STM signatures [compare with Figs.~\ref{Fig1}(a) and~\ref{Fig1}(b)]. This allows us to unambiguously assign the observed Fe$_{\rm{A}}$ and Fe$_{\rm{B}}$ species to Fe adatoms in hcp and fcc positions, respectively.

The observation of single peaks in the STS measurements [Fig.~\ref{Fig1}(c)] suggests that Fe adatoms on the surface of Bi$_2$Te$_3$ behave as simple scalar impurities \cite{Biswas2010}. In other words, their magnetic moments show only a weak coupling to the topologically protected surface states. The comparison with the results of our first-principles calculations is not conclusive since the computed density-of-states peaks show additional splittings as a result of the supercell approximation employed in our models (not shown here). Besides these structural and electronic properties of the adatoms, we used the LDA+$U$ calculations to estimate the magnetic properties of the adatoms. The computed spin and orbital magnetic moments for the fcc-site Fe adatom with an out-of-plane spin orientation are $m^{S}_{\rm fcc}=2.7~\mu_{\text{B}}$/atom and $m^{L}_{\rm fcc}=0.7~\mu_{\text{B}}$/atom, respectively. For the hcp-site adatom we find $m^{S}_{\rm hcp}=2.5~\mu_{\text{B}}$/atom and $m^{L}_{\rm hcp}=0.3~\mu_{\text{B}}$/atom. Furthermore, we find strong out-of-plane anisotropies for both species, i.e. $K_{\rm fcc}=12$~meV and $K_{\rm hcp}=10$~meV, where $K$ denotes the magnetocrystalline anisotropy energy per atom. This prediction is contrary to a recent experimental report where an easy-plane anisotropy of Fe/Bi$_2$Te$_3$ was assumed based on orbital symmetry arguments~\cite{Shelford2012}.

\begin{figure}[b]
\begin{center}
\includegraphics[width=0.48\textwidth]{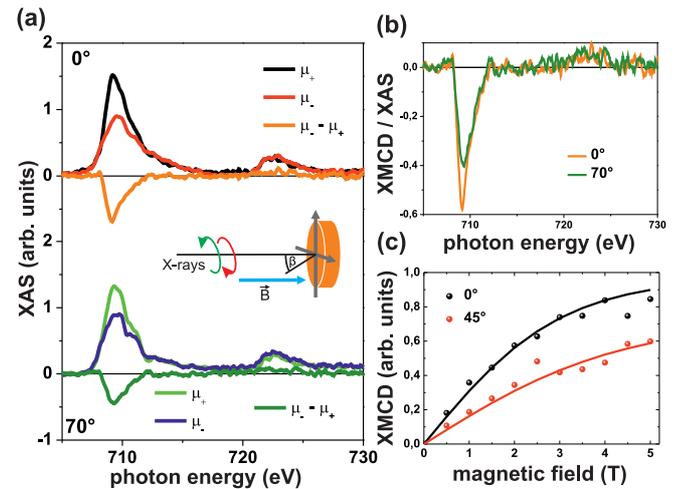}
\caption{(Color online) (a) Circularly polarized x-ray absorption and XMCD spectra of Fe adatoms on Bi$_2$Te$_3$ measured at the Fe $L_{3/2}$ edges at a temperature of 6~K. The upper (lower) panel reveals data obtained by 0.01 MLE Fe for a normal (grazing) incidence angle. The inset shows a sketch of the experimental setup. (b) XMCD signals measured at $B=5$~T divided by the XAS $L_3$-edge intensity, indicating an out-of-plane easy axis. (c) Element-selective magnetization curves for the Fe adatoms at varying angles. The data points represent the Fe XMCD intensity measured at the $L_3$ edge while the solid lines depict the best-fit profiles. The $y$ axis has been scaled relative to the saturation magnetization determined by the fitting procedure.}
\label{Fig3}
\end{center}
\end{figure}

To investigate the electronic and especially the magnetic properties of Fe on Bi$_2$Te$_3$ in more detail, we performed XAS and XMCD measurements, which are summarized in Fig.~\ref{Fig3}. The XAS spectra of 0.01 MLE Fe/Bi$_2$Te$_3$ reveal a faint multiplet structure at the $L_3$ edge for a normal incidence angle [Fig.~\ref{Fig3}(a), upper panel] and a grazing  [Fig.~\ref{Fig3}(a), lower panel]. The spectral shape suggests a ground state with an electronic configuration of $3d^6$ of the ensemble of Fe adatoms~\cite{Groot1990,Laan1992} which is consistent with a previous report on the same sample system~\cite{Shelford2012}. From the integrated total XAS we find a branching ratio~\cite{branching,Thole1988} of $BR=0.85 \pm 0.01$, suggesting a high-spin ground state of the Fe adatoms. Furthermore, the normalized dichroism signal strength, i.e., XMCD divided by the $L_3$ XAS intensity, can be used as an indicator for the anisotropy of the system. We find this quantity to be enhanced by about 50\% for a normal incidence angle, indicating an out-of-plane easy axis of the Fe adatoms [Fig.~\ref{Fig3}(b)], which is in agreement with our predictions and contrary to Ref.~\onlinecite{Shelford2012}. We confirmed the out-of-plane easy axis by means of magnetization curves, revealing paramagnetic behavior of the Fe magnetic moments and showing the steepest slope at 0$^{\circ}$ [Fig.~\ref{Fig3}(c)].

To analyze the magnetization curves, we fitted the adatom magnetization in the framework of a classical thermodynamical model. This model is based on a magnetic Hamiltonian including the Zeeman splitting as well as the anisotropy energy and a Boltzmann term~\cite{Brune2009}. To generalize the model even more, we distinguished the contributions of the Fe adatoms in both positions by setting them to 60\% and 40\% in advance of fcc-site adatoms in accordance with the STM observations. Due to the intrinsically ill-conditioned fit problem we used some information from our calculations as fit constraints; namely, we assumed total magnetic moments of $m_{\text{fcc}}^{\text{fit}} = 3.4$~$\mu_{\rm B}$/atom and $m_{\text{hcp}}^{\text{fit}} = 2.8$~$\mu_{\rm B}$/atom. Hence, two main fit parameters are defined by the contributions' magnetic anisotropies given in meV per atom, at which a ratio of $K_{\text{fcc}} / K_{\text{hcp}} = 1.2$ has been assumed. Further parameters are the temperature $T$ (known from the experiment) and the saturation magnetization $m_{\rm sat}$. The best fit is obtained for $K_{\text{fcc}} = (10 \pm 4)$~meV/atom and $K_{\text{hcp}} = (8 \pm 4)$~meV/atom. The size of the errors is mainly given by the ``scattered'' character of the measured XMCD intensities. Within the error bars the fit results agree well with our theoretical predictions.

In order to verify this hypothesis, the XAS/XMCD spectra acquired at normal incidence angle were analyzed by means of the sum rules~\cite{Thole1992, Carra1993}. Although from Fig.~\ref{Fig3}(c) it is obvious that the external magnet was not sufficiently strong to fully saturate the magnetic moments of the sample, the curves are almost flat at the maximum field applied ($B = 5$~T). The saturation magnetization found by our fitting procedure reveals that the magnetization reaches about 90\% of the expected saturation value at normal incidence for 5 T. Hence, we evaluated the orbital moment ($m^{L}$) and the effective spin moment ($m^{S\text{, }eff}=m^S+7m^D$) as well as the ratio of orbital to effective spin moment~\cite{ratio}. We obtained $m^{L}=(1.1 \pm 0.1)~\mu_{\text{B}}$/atom, $m^{S\text{, }eff}=(2.9 \pm 0.2)~\mu_{\text{B}}$/atom, and $R=0.38 \pm 0.04$. Note, that $R$ is about eight times larger than the bulk value of Fe~\cite{Chen1995}, indicating the character of individual Fe adatoms investigated here~\cite{Gambardella2002}. On the one hand, the magnetic moments were not fully saturated. For this reason, the values of $m^{L}$ and $m^{S\text{, }eff}$ have to be normalized by a factor of $0.9^{-1}$. On the other hand, determining the pure spin moment requires separating the contribution of the spin dipole moment. The latter describes an anisotropy of the spin density if the atomic cloud is distorted either due to spin-orbit interaction or crystal field effects. In a previous work~\cite{Crocombette1996}, Crocombette \textit{et al.} calculated the contribution of the spin dipole moment for Fe adatoms on a surface at low temperature to be $7 m^D = 0.25 \cdot m^S$, which is in agreement with later investigations on the applicability of the XMCD sum rules~\cite{Brune2009, Piamonteze2009, Sipr2009}. We hence estimate the pure spin moment to be $m^S= m^{S\text{, }eff} \cdot 0.9^{-1} \cdot 1.25^{-1} = (2.6 \pm 0.2)~\mu_{\text{B}}$/atom. Comparing the values of $m^L$ and $m^S$ determined by the sum rules to the values of the LDA+$U$ approach reveals perfect agreement in the case of the spin moment. The orbital moments were calculated from the projection of the Kohn-Sham wave functions on Fe atomic orbitals. This approximation is justified for isolated magnetic adatoms but only contains the contribution of the Fe orbitals and thus underestimates the experimental value. Nevertheless, the overall increase of the orbital moment with respect to the bulk value is well captured by the calculation. A summary of the experimentally obtained as well as the theoretically predicted magnetic moments and anisotropies is shown in Table~\ref{table1}.

\begin{table}
\begin{ruledtabular}
\caption{Summary of the magnetic moments and anisotropies of Fe/Bi$_2$Te$_3$ experimentally determined by the sum rules and the magnetization curve fitting procedure as well as calculated using the LDA+$U$ approach. Details are given in the text.} 
\begin{tabular}{lccc}
\multirow{2}*{Method} & $m^{S}$ & $m^{L}$ & $K$ \\
 & ($\mu_{\text{B}}$/atom) & ($\mu_{\text{B}}$/atom) & (meV/atom) \\
\addlinespace[1.5pt]
\hline
\addlinespace[1.5pt]
\multirow{2}*{LDA+$U$} & 2.7 (fcc)  & 0.7 (fcc) & 12 (fcc) \\
& 2.5 (hcp)  & 0.3 (hcp) & 10 (hcp) \\
\addlinespace[1.5pt]
\hline
\addlinespace[1.5pt]
Sum rules & 2.6~$\pm$~0.2 &  1.2~$\pm$~0.1 & --\\
\addlinespace[1.5pt]
\hline
\addlinespace[1.5pt]
magnetization & & & ($10 \pm 4$) (fcc)\\
curve fitting & & & ($8 \pm 4$) (hcp)\\
\end{tabular}
\label{table1}
\end{ruledtabular}
\end{table}

In the following, the observations made here about individual Fe adatoms on the (111) surface of Bi$_2$Te$_3$ are discussed in view of already published experimental studies dealing with Fe deposition on TIs. Note that this type of surface doping significantly differs from bulk-doping where Fe is introduced already in the crystal growth. For this reason, studies based on bulk doping effects are not discussed here.

By means of ARPES experiments, Wray \textit{et al.}~\cite{Wray2011} observed a significant $n$-doping effect due to Fe deposition on the surface of Bi$_2$Se$_3$. Moreover, for a coverage of about 0.5 MLE, an opening of a gap at the DP was observed. The $n$-doping effect of Fe impurities is well in line with several experimental studies on Bi$_2$Se$_3$~\cite{Scholz2012,Honolka2012,Schlenk2013,Ye2013} and with the observations made here. Interestingly, Scholz \textit{et al.}~\cite{Scholz2012} observed a $p$-doping effect if the sample was cooled down to 8~K after room-temperature deposition of Fe. However, the initial deposition of Fe causes an $n$-doping effect. An opening of a gap at the DP was not observed within these studies~\cite{Scholz2012,Honolka2012,Schlenk2013,Ye2013}.

After low-temperature deposition of Fe/Bi$_2$Se$_3$, Honolka \textit{et al.}~\cite{Honolka2012} revealed the existence of two different species of Fe adsorbates on the surface. These were identified to be adsorbed in the fcc and hcp hollow sites of the surface Se layer. By means of XMCD and element-specific magnetization curves an easy-plane anisotropy of the Fe magnetic moments was found. These findings are in line with DFT calculations utilizing the general gradient approximation if dynamical hybridization effects are included~\cite{Honolka2012}. The easy-plane anisotropy of the Fe magnetic moments is contrary to the results obtained by Ye \textit{et al.}~\cite{Ye2013}. Importantly, the latter work is based on room temperature deposition of Fe. In this case, the available thermal energy may lead to clustering of the Fe impurities which presumably modifies the electronic properties of the adsorbates significantly. For this reason, the different preparation conditions might be the cause for the contrasting results. Moreover, Schlenk \textit{et al.}~\cite{Schlenk2013} showed that room-temperature treatment of Fe impurities on Bi$_2$Se$_3$ causes an additional effect. After low-temperature deposition of Fe a warming of the samples induced a segregation process of Fe atoms into the bulk crystal already at 260~K. There, the Fe atoms substituted Bi atoms at their lattice sites. As a consequence, dark triangular features are visible in STM topographies well known from bulk-doped crystals~\cite{Song2012}. This finding additionally stresses the importance of the preparation conditions when different studies are compared to each other.

For the particular system of Fe impurities on Bi$_2$Te$_3$ literature is rare. As already mentioned above, West \textit{et al.}~\cite{West2012} investigated Fe/Bi$_2$Te$_3$ by means of STM. After the deposition at 50~K they observed a single species of Fe adsorbates contrary to our observations. In view of the discussion related to Bi$_2$Se$_3$, the different preparation conditions reflect a likely cause for the different experimental results. Shelford \textit{et al.}~\cite{Shelford2012} investigated this system by means of XMCD after low-temperature deposition (1.5~K) of Fe. In agreement with our results no ferromagnetic order was detected. Moreover, they estimated the ratio of the orbital magnetic moment to the effective spin moment to be 0.46, which indicates a considerably large orbital moment. This finding particularly supports the orbital magnetic moment obtained from the XMCD sum rules in the study at hand. According to Table~\ref{table1}, the ratio reads $R_{\text{sum rules}}=0.38 \pm 0.04$. It thus deviates only slightly from that reported before. Based on orbital symmetry arguments, Shelford \textit{et al.}~\cite{Shelford2012} concluded that Fe/Bi$_2$Te$_3$ is expected to reveal an easy-plane anisotropy because the $d_{3z^2-r^2}$ orbital is lowest in energy and has no out-of-plane orbital moment. Although the remaining observations are quite similar to our results, this conclusion contradicts our experimental findings.

In view of the work by Schlenk \textit{et al.}~\cite{Schlenk2013}, similar processes, namely, a bulk segregation of surface impurities, should be considered also for Bi$_2$Te$_3$ if experimental studies are performed at similar temperatures. As a consequence, for aiming toward a basic understanding of the system of Fe/Bi$_2$Te$_3$ low temperatures should be chosen. For the particular case of low-temperature deposition of Fe, we experimentally verify an out-of-plane anisotropy of the Fe magnetic moment contrasting with that of Fe/Bi$_2$Se$_3$~\cite{Honolka2012}. The difference may be caused by structural differences among both surfaces, which are suggested to lead to a different hybridization with the substrate thereby causing modified electronic and magnetic properties.

In summary, our results provide deep insight into the magnetic and electronic properties of Fe adatoms on the surface of the topological insulator Bi$_2$Te$_3$. For low-temperature deposition we identify two different species of Fe adatoms adsorbed in the fcc and hcp hollow sites. The magnetic moments of Fe show paramagnetic behavior while their electronic configuration of the ground state is $3d^6$. Notably, the relative magnitude of the normalized XMCD spectra and the magnetization curves confirm the theoretical prediction of very large out-of-plane anisotropies for this system.

Financial support from the ERC Advanced Grant ``FURORE'' and the German Research Foundation (DFG) in the framework of the SFB 668 is gratefully acknowledged. A.P., G.A., and O.V.Y. were supported by the ERC Starting Grant ``TopoMat'' (Grant No. 306504) and by the Swiss National Science Foundation (Grant No. PP00P2\_133552). Computations have been performed at the Swiss National Supercomputing Centre (CSCS) under Project No. s443. Fitting was performed using the PL-Grid Infrastructure. M.S. acknowledges the support of the Polish Ministry of Science and Higher Education.


\begin{thebibliography}{48}
\bibitem [{rat()}]{ratio}%
  \BibitemOpen
  \href@noop {} {}\bibinfo {howpublished} {$R$: Ratio of the orbital magnetic
  moment ($m^L$) to the sum of the spin moment ($m^S$) and the spin dipole moment
  ($m^D$) according to $R$~=~$m^L$~/~($m^S$~+~7~$m^D$), where the moments are
  given in $\mu_{\rm B}$/atom and are projected along the incident beam
  direction.}\BibitemShut {Stop}%
\end{thebibliography}
\end{document}